\documentclass[useAMS,usenatbib]{mn2e}

\usepackage{times}
\usepackage{graphicx}
\usepackage{amsmath,amssymb,amsfonts} 
\usepackage{multirow}
\usepackage[utf8]{inputenc}
\usepackage{ctable}

\usepackage{lipsum}
\usepackage{color}

\def\nh{$N_{\rm H}$}

\def\msun{M$_{\odot}$}

\def\gx339{\mbox{GX{339--4}}}
\def\h1743{\mbox{H1743--322}}
\def\cygx{\mbox{Cyg\,~X-1}}

\newcommand{\xrb}{\mbox{X-ray} binary}
\newcommand{\xrbs}{\mbox{X-ray} binaries}
\newcommand{\sed}{spectral energy distribution}

\def\rxte{\textit{Rossi X-ray Timing Explorer}}

\newcommand{\jwst}{\textit{James Webb Space Telescope}}
\newcommand{\integral}{\textit{INTEGRAL}}

\newcommand{\ishem}{\texttt{ishem}}

\title[Dark jets in the soft X-ray state of black hole binaries?]
{Dark jets in the soft X-ray state of black hole binaries?}
\author[S.Drappeau et al.]{S.Drappeau,$^{1,2}$\thanks{E-mail: samia.drappeau@irap.omp.eu}
    J.Malzac,$^{1,2}$
    M.Coriat,$^{1,2}$
    J.Rodriguez,$^{3}$
    \newauthor
    T.M.Belloni,$^{7}$
    R.Belmont,$^{1,2,3}$
    M.Clavel,$^{3,4}$
    S.Chakravorty,$^{5,6,9}$
    S.Corbel,$^{3}$
    J.Ferreira,$^{5,6}$
    \newauthor
    P.Gandhi,$^{8}$
    G.Henri$^{5,6}$
    and P.-O.Petrucci,$^{5,6}$\\
 $^{1}$Universit\'{e} de Toulouse; UPS-OMP; IRAP; Toulouse, France\\
 $^{2}$CNRS; IRAP; 9 Av. colonel Roche, BP 44346, F-31028 Toulouse cedex 4,
 France\\
 $^{3}$Laboratoire AIM (CEA/IRFU - CNRS/INSU - Universit\'{e} Paris Diderot), CEA
 DRF/IRFU/SAp, F-91191 Gif-sur-Yvette, France\\
 $^{4}$ Space Sciences Laboratory, University of California, Berkeley CA 94720, USA\\
 $^{5}$Universit\'{e} Grenoble Alpes, IPAG F-38000 Grenoble, France\\
 $^{6}$CNRS, IPAG, F-38000 Grenoble, France\\
 $^{7}$Osservatorio Astronomico di Brera, Istituto Nazionale di Astrofisica, Via E. Bianchi 46, I-23807, Merate, Italy\\
 $^{8}$Department of Physics \& Astronomy, University of Southampton, Highfield, Southampton SO17 1BJ\\
 $^{9}$Department of Physics, Indian Institute of Science, Bangalore, 560012, India\\
}

\begin{document}

\date{Accepted 2016 December 14. Received 2016 December 13; in original form 2016 April 14}

\pagerange{\pageref{firstpage}--\pageref{lastpage}} \pubyear{2016}

\maketitle

\label{firstpage}

\begin{abstract}
 \xrb\ observations led to the interpretation that powerful compact jets,
 produced in the hard state, are quenched when the source transitions to its
 soft state. The aim of this paper is to discuss the possibility that a
 powerful dark jet is still present in the soft state. Using the black
 hole X-ray binaries \gx339\ and \h1743\ as test cases, we feed observed
 X-ray power density spectra in the soft state of these two sources to an
 internal shock jet model. Remarkably, the predicted radio emission is
 consistent with current upper-limits. Our results show that, for these two
 sources, a compact dark jet could persist in the soft state with no major
 modification of its kinetic power compared to the hard state.
\end{abstract}

\begin{keywords}
accretion, accretion discs -- black hole physics -- shock waves -- relativistic
processes -- radiation mechanisms: non-thermal -- X-rays: binaries
\end{keywords}

\section{Introduction}
The properties of outflows launched by accreting black holes in X-ray binary
systems appear to be deeply connected to the state of the accretion flow
\citep[see e.g.][]{Doneetal2007, FenderGallo14, Malzac2016}. In their hard X-ray
spectral state, black hole \xrbs\ emit powerful compact jets. Those jets radiate
partially self-absorbed synchrotron emission, which is routinely detected in the
radio band \citep{Fenderetal2000, Fender2001} with flat or weakly inverted
spectral slopes ($F_\nu\propto \nu^{\alpha}$ with $\alpha\simeq-0.5\,$-$\,0$).
This emission can extend at higher frequencies up to the IR and optical bands
\citep{CorbelFender2002, Chatyetal2012, Gandhietal2011}. In the
prototypical source \cygx\ , GeV emission was recently detected by Fermi
\citep{Malyshevetal2013, Zaninetal2016} and interpreted as inverse Compton emission
from the jet \citep{Zdziarskietal2014, Zdziarskietal2016}. Also, in this source, the strong
polarization fraction recently detected by \integral\  above 400keV
\citep{Laurentetal2011, jourdainpol12, Rodriguezetal2015} suggests that the MeV tail
observed in the hard state  \citep{McConnelletal2000, Jourdainetal2012, Zdziarskietal2012}
originates from optically thin synchrotron emission in the jet.

In some cases, radio jet structures have been resolved
at the AU scale \citep{Stirlingetal2001,Fuchsetal2003}.  Observations suggest
that the jet kinetic power in the hard state could be comparable to, or even
larger than, the X-ray luminosity \citep{Galloetal2003, Galloetal2005,
Kordingetal2006a}. In contrast, compact jets are not detected in the soft state.
The current radio upper-limits indicate that the emission is suppressed by
several orders of magnitude at least \citep{Corbeletal2000, Russelletal2011}.
This led to the generally accepted view that the jets are not produced in the
soft state \citep{Fenderetal2004, Fenderetal2009}.

Here we will argue that
compact jets could still be present in the soft state, with a kinetic power
comparable to that in the hard state, and that only their emission is quenched.
This is an important issue also because this would affect the estimates of the
global kinetic feedback of accreting black holes with consequences for the large
scale impact of supermassive black holes on their environment. 

A popular model for the radio-IR emission of black hole jets is the
\textit{internal shock model}. \citep{Rees1978a, ReesMeszaros1994,
DaigneMochkovitch1998, KaiserSunyaevSpruit2000, Spadaetal2001,
BottcherDermer2010, JamilFenderKaiser2010, Malzac2013}. In this model,
velocity fluctuations injected at the base of the jet drive internal shocks at
large distances from the black hole. Leptons are accelerated at the shocks and
emit synchrotron radiation. Malzac \citeyearpar{Malzac2013, Malzac2014} showed
that the total radiated power and the \sed\ (SED) are very
sensitive to the amplitude and time-scales of the velocity fluctuations. The
origin of the velocity fluctuations is not specified in the model but are likely
to be driven by the fluctuations of the accretion flow which in turn can be
traced by the X-ray light curve. If this is the case, the jet velocity
fluctuations are expected to be similar to the observed X-ray fluctuations.
Therefore assuming the jet Lorentz factor variation have the same Fourier power
density spectrum (PDS) as the observed X-ray PDS, the model should predict a
radio-IR jet \sed\ that is close to the observations. The results of
\citet[][hereafter D15]{Drappeauetal2015} suggest that this is indeed the case.
They used one of the most complete multi-wavelength observation of \gx339 in
the hard state and successfully modelled the \mbox{radio-IR} SED using the observed
X-ray PDS as input of the model.

The typical rms amplitude of the fast ($\la$ 1 ks) X-ray variability is in the
range 10 -- 30 percent in the hard state, and decreases to less than a few percent
in the soft state \citep[see e.g.][]{BelloniStella2014}. In the framework of
the internal shock model, smaller amplitude fluctuations of the jet Lorentz
factor imply weaker shocks and less energy available to particle acceleration
and radiation. The resulting radio flux scales approximately like $\sigma^{2.8}$,
where $\sigma$ is the fractional rms amplitude of fluctuations \citep[see][equation~39]{Malzac2013}.
Therefore, if the amplitude of the jet fluctuations tracks the X-ray variability
also in the soft state, the radio emission is expected to drop by several orders
of magnitude with respect to the hard state level, even if the jet kinetic power
remains unchanged. 

In this paper we investigate whether this effect could be sufficient to explain
the observed disappearance of the radio emission in the soft state. We model the
radio emission of two black hole binaries \gx339\ and \h1743 in the soft state
using the internal shock model \ishem\ \citep{Malzac2014} and compare the
results to the observational upper-limits. The data and model are described in
Sec.~\ref{sec:obs} and Sec.~\ref{sec:meth} respectively. We find that the weak
fluctuations observed in the soft state produce weak radio emission compatible
with the current upper-limits, whereas the total kinetic power of such
\textit{dark} jets can be as large as in the hard state. These findings are
discussed in Sect.~\ref{sec:results} and \ref{sec:discussion}.

\section{Observations and data reduction}\label{sec:obs}
For both \gx339\ and \h1743\ we use X-ray PDS obtained from soft state
observations. In the case of \h1743\ we focused on a single observation that
was obtained in (quasi-)simultaneity with a radio observation. The \rxte\
(\textit{RXTE}) observations was made on 2003 august 11. For \gx339, since no
simultaneous radio - X-ray observations were available, we accumulated all the
data obtained during the soft state of the 2010 outburst \citep[from 2010 May 14
to December 30; see e.g.][]{Claveletal2016}. This allowed us to increase the
statistical significance of the PDS obtained. 

For both sources we produced $\sim 4$~ms ($2^{-8}$~s) resolution light curves
from the Proportional Counter Array (PCA) onboard \textit{RXTE}. The data
were reduced in a standard manner
\citep[e.g.][]{RodriguezandVarniere2011,Claveletal2016} with the
{\tt{HEASOFT}} V6.16 software suite. The light curves were obtained from event
mode data between $\sim 2$ and $\sim 50$~keV (spectral channels 0 -- 116) in
order to limit the background noise above this energy. We also checked that no
differences in the final result were obtained while limiting further the energy
range to $\sim$ 2 -- 20 keV. PDS from each individual observation were produced
on intervals of 96~s, all intervals were further averaged, and, in the case of
\gx339, all observations were combined to produce a single PDS. The dead-time
corrected white noise was subtracted from the PDS. The resulting soft state PDS
of \gx339\ and \h1743\ are shown in Figure~\ref{fig:pds}, together with the hard
state PDS of \gx339\ used in D15. The 0.07--5 Hz fractional rms amplitudes are
respectively 1.9, 2.3, and 27 percent.

Radio observations of \gx339\ used in this study were conducted with the
Australia Telescope Compact Array (ATCA) on 2010 June 25 and 2010 August 22 when
\gx339\ was in the soft state. The array was in the extended 6C and compact H168
configurations during the June and August observations respectively. We observe
the source at 5.5~GHz and 9~GHz simultaneously. Each frequency band was composed
of 2048 1-MHz channels. For both observations, we used PKS B1934-638 for
absolute flux and calibration and PKS J1646-50 to calibrate the antenna gains as
a function of time. Flagging, calibration and imaging were carried out with the
Multi-channel Image Reconstruction, Image Analysis and Display software
\citep[MIRIAD,][]{Saultetal1995}. We did not detect the source in any of the
observations. To obtain the best constraint on jet emission of \gx339, we
combined the two observations at imaging step and obtain a 3$\sigma$ upper-limit
of 24 $\mu$Jy. The optical and near-infrared (hereafter OIR) observations were
taken with the ANDICAM camera on the SMARTS 1.3 m telescope located at Cerro
Tololo in Chile \citep[see][]{Buxtonetal2012}. The values used for this
study are an average of the different measurements taken between the two radio
observations. The error is estimated from the variance of the flux over that period.

The radio 8.46~GHz upper-limit of 30~$\mu$Jy used for the study of \h1743 have
been published by \citet{McClintocketal2009} based on a VLA observation
conducted on MJD 52863 during the soft state of its 2003 outburst. The OIR
observations have been published by \citet{Chatyetal2015}. We refer the
reader to these articles for further details. 

\section{Jet emission model}\label{sec:meth}
\subsection{\ishem\ model}
We use the numerical code \ishem\ which simulates the hierarchical merging and
the emission of ejecta constituting a jet. At each time step $\Delta t$,
a new shell of matter is ejected. $\Delta t$ is comparable to the dynamical
timescale at the initial ejecta radius $r_{dyn}$. Each new created shell has
a specific Lorentz factor $\gamma$.  Its value depends on the time of the
ejection, so that the overall distribution of its fluctuations follows the
shape of a given PDS. All shells, the one injected and the one resulting
from mergers, are tracked throughout the duration of the simulation until
they merge with other shells. When propagating outwards, adiabatic losses
cause the shells to gradually loose their internal energy. Eventually, while
merging, part of their bulk kinetic energy is converted into internal energy and
radiation. The details of the physics and the description of the main
parameters of the model are presented in the original paper \citep{Malzac2014}.

\subsection{Input parameters from \gx339\ and \h1743}
The low mass black hole binary \gx339\ is a recurrent X-ray transient. The
source properties are not well constrained. Here we assume a black hole mass of
10~\msun, a distance of 8 kpc, and an inclination angle of $23^\circ$ compared
to the line of sight.  These parameters are within the observational
uncertainties \citep{Zdziarskietal2004, Shidatsuetal2011} and identical to those
used in D15.  \h1743\ is an X-ray binary and a black hole candidate, exhibiting
recurrent outbursts. We assume a mass of 10~\msun, an inclination angle of
$75^\circ$ and a distance of 8.5 kpc \citep{Steineretal2012}. For both sources,
the jet half-opening angle $\phi$ is set to $1^\circ$ (unless otherwise stated)
and the time-averaged jet Lorentz factor is set to $\langle\Gamma\rangle=2$ in
agreement with current observational constraints \citep{Fenderetal2009}.

An important parameter of the model is the kinetic power available to the jet.
In the case of \gx339\, we choose the value estimated in
D15 to be able to perform comparison with the study done
in the hard state for this source, $P_{jet} \simeq 0.05 \, L_{Edd}$. As for
\h1743, the total power of the jets is set to equal the observed X-ray
luminosity i.e. $P_{jet} = 0.06 \, L_{Edd}$. Note that $P_{jet}$ is the total
kinetic power of the two-sided jets. Only a fraction (potentially very small) of
this total kinetic power can be radiated away.\footnote{In the case of blazar
jets this radiated fraction may reach 10 percent of their kinetic power
\citep{Ghisellinietal2014}, which is achievable in the colliding shell model if the
amplitude of the velocity fluctuations is large enough.  In the framework of
gamma-ray bursts, \citet{Beloborodov2000} finds that the radiative efficiency of
internal shocks can reach almost 100 percent.}

\begin{figure}
 \centering
\includegraphics[width=0.45\textwidth]{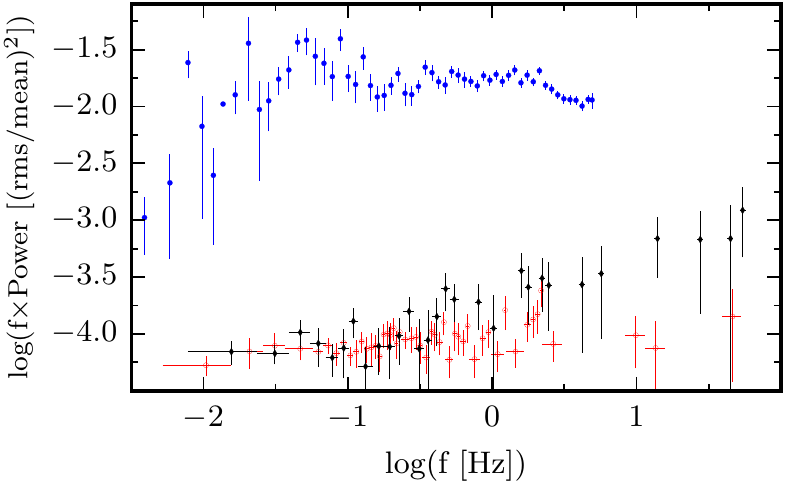}
\caption[X-ray PDS]{Soft state X-ray PDS of \gx339\ (red circle) and \h1743\
(black diamond) in the \mbox{2 -- 50 keV} band, used to constrain the
fluctuations of the bulk Lorentz factor of the ejecta. For comparison, the X-ray
PDS of \gx339\ in the hard state is also shown (blue filled circle).}
\label{fig:pds}
\end{figure}

Besides these inputs, the time-distribution of the fluctuations of
the kinetic energy $(\gamma - 1)$ has a power spectrum which follow the shape
(and amplitude) of the observed X-ray PDS, for both sources, in the soft state
(see Fig.~\ref{fig:pds}). We extrapolate the shape at low frequencies (down to
$10^{-5}$ Hz) and high frequencies (up to $50$ Hz), assuming in both
case $PDS \propto \frac{1}{f}$.

\subsection{Simulation and outputs}
To produce the emission of fully developed jet and counter-jet, we run our
simulations for a time $t_{simu} = 10^5\,\mathrm{s}$ \mbox{($\sim 1$ day)},
as measured in the observer's frame.
Self-absorbed synchrotron from a non-thermal population of electrons is the only
radiation process considered here. The electrons have a power-law distribution
with a spectral index of $p = 2.3$, between minimum energy $\gamma_{min}=1$ and
maximum energy $\gamma_{max}=5 \times 10^{3}$. These values are set and fixed
throughout the simulation. The choice of $p=2.3$ is consistent with the typical
value expected in shock acceleration. These parameters are identical to
those used in D15 except for  $\gamma_{max}$ which was reduced (from $10^6$ to
$5\times10^3$) to suppress any contribution of the jet emission in the
\textit{RXTE} band (see discussion in Section~\ref{sec:results}).

The emission from each individual shell created
during the simulation is time-averaged over the simulation running time
$t_{simu}$ to produce the final SED. The general shape of the simulated SED is
determined solely by the shape of the PDS we used. The other free parameters of
the model only allow us to modify the flux normalisation or to shift it with
respect to the photon frequency. The broad-band spectra are computed from $10^7$
to $10^{16}$ Hz.

To assess the plausibility of our model, we compare the simulated
SEDs to the radio upper-limits obtained from the observations (see
Section~\ref{sec:obs}).

\begin{table*}
\caption{Parameters of the \textsc{diskir} models shown in Fig.~\ref{fig:sed_gx339} and Fig.~\ref{fig:sed_h1743}, the inner irradiation fraction, $f_{\rm in}$, is fixed to 0.1 in all models. The galactic absorption column \nh\ was fixed to 0.4 and 2 $\times10^{22}\,{\rm cm}^{-2}$  in \gx339\ and \h1743\ respectively.}
\label{tab:diskir}
\begin{tabular}{lcccccccccc}
\hline
                               &     $kT_{\rm disc}$ (keV)  &     $\Gamma$      &  $kT_{\rm e}$ (keV)  &  $L_{\rm c}/L_{\rm d}$       &     $f_{\rm out}$    &    $R_{\rm irr}/R_{\rm in}$      &    ${\rm log}\frac{R_{\rm out}}{R_{\rm in}}$ &  Normalization\\
  \hline
   \gx339\ (hard)  &       0.19                         &     1.61                  &   44                           &  4.6      				  & 0.04           &   2                        &   3.8  & $10^5$   \\
   \gx339\ (soft)   &       0.80                          &    2.34                  &    915                         &   0.14                                &$1.6 \times 10^{-2}$  &   1.01               &    4.8   &  2693 \\ 
    \h1743\      (soft)  &       0.98      &     2.22     &       200               & $0.014$  &   $7.92\times 10^{-4}$ 					&    10       &     4.1   &  1120\\
\hline
\end{tabular}
 \end{table*}

\section{Results}\label{sec:results}
\begin{figure*}
 \centering
 \begin{minipage}{180mm}
 \includegraphics[width=0.95\textwidth]{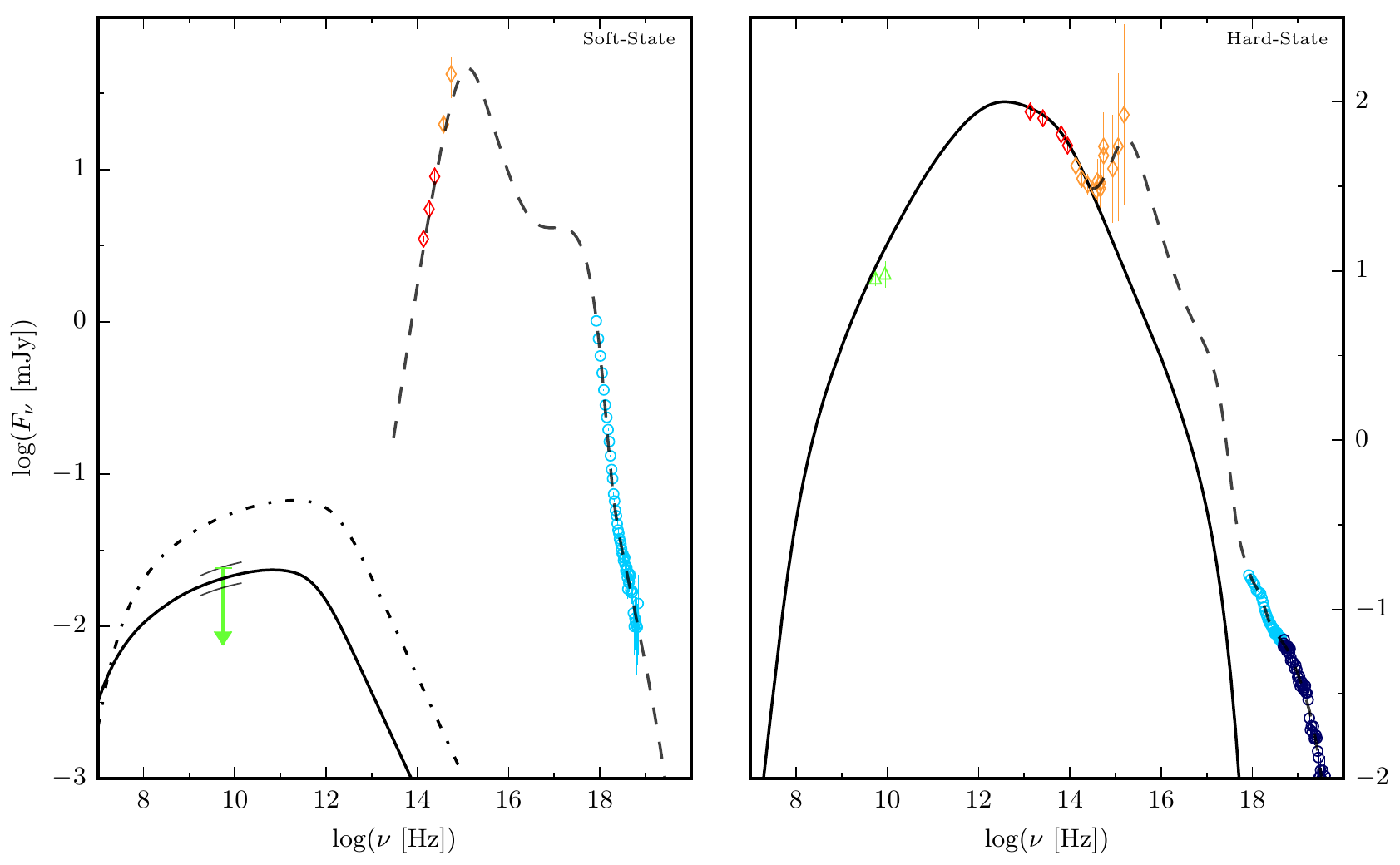}
 \caption[SED]{Broadband spectra of \gx339\ in the soft state (left)
 compared to that of the hard state (right), studied in
 D15. Colour code of the observed data is as follow: In green are the Radio
 observations and upper-limits, in red are the infrared bands, in orange the optical and
 ultraviolet, and in blue the X-ray (light blue is PCA,
     dark blue is HEXTE. Soft-state PCA data were observed on May 19,
     2010). The vertical error bars represent the statistical and
 systematic errors on the mean. On the left panel, the total self-absorbed
 synchrotron jet emission from our two models ($\phi = 3^\circ$ and $\phi =
 1^\circ$) are shown as solid and dot-dash black lines, respectively. The two
 thin black curves around the radio upper-limit represent the radio flux predicted from a $2.5^\circ$ and a
 $3.5^\circ$ half-opened jet. The spectra have been averaged over the whole
 duration of the simulation. The OIR to X-ray emission is assumed to
 originate from the accretion flow and the data are fit  with the irradiated
     accretion disc \textsc{diskir} model, shown in dashed lines in both
     panels. We note that the
 right panel corrects Figure 2 of D15 in which the \textit{RXTE} data were erroneously
 plotted with a normalisation that was too large by a factor of 1.5. Also
differently than the model SED of D15, the jet model now also includes a synchrotron
 cut-off associated to maximum energy of the electrons (see discussion in
 section~\ref{sec:results}).}
 \label{fig:sed_gx339}
 \end{minipage}
\end{figure*}

Figure~\ref{fig:sed_gx339} (left) compares a simulated SED to the observed SED
of \gx339\ in the soft state. The simulated and observed SEDs in the hard state
of D15 are also shown for comparison on the right.
The soft state dash-dot SED
show the result of a simulation in which all our model parameters for the soft
state are identical to those of the hard state, except for the shape of the
fluctuations of the jet bulk Lorentz factor. The significant difference between
the hard and soft state simulated SEDs is then only due to the different PDS.
We see that compared to the hard state, the predicted radio flux (shown in
dash-dot lines) drops by almost three orders of magnitude in the soft state.
However, it is right above the observational upper-limits and therefore should
have been detected. Nevertheless, slightly widening the jet half-opening angle
to $3^\circ$ drops the radio flux by half an order of magnitude, below the
observational upper-limit. This simulation is shown by the thick black curve on
the left panel of Fig.~\ref{fig:sed_gx339}. The predicted radio fluxes for a
$2.5^\circ$ and a $3.5^\circ$ half-opening angle are also shown.
For larger opening angles, the same amount of dissipation occurs in a
larger jet volume. As a result the particles energy density and magnetic field
are reduced which in turn reduces the synchrotron emission.

\begin{figure}
 \centering
 \includegraphics[width=0.45\textwidth]{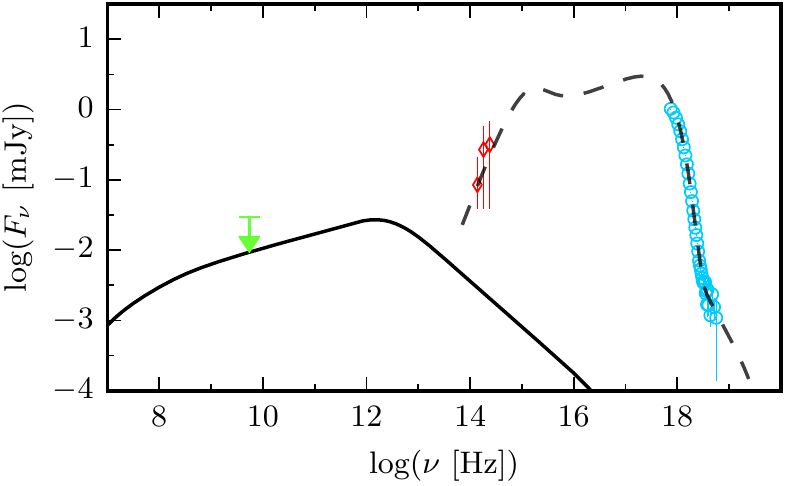}
 \caption[SED]{Broadband spectra of \h1743\
 in the soft state. The radio upper-limit is plotted in green, and the
 near-infrared and the X-ray observations are in red and in blue, respectively.
 The vertical error bars represent the statistical and systematic errors on the mean.
 The total self-absorbed synchrotron jet emission from our model, averaged over
 the whole duration of the simulation, is shown as solid black. X-ray emission
 is used here solely as upper-limits to define the feasibility of our fit.
 Its emission is fitted with the irradiated accretion disc \textsc{diskir} model, shown in dashed line.}
 \label{fig:sed_h1743}
\end{figure}

Figure~\ref{fig:sed_h1743} shows the simulated SED of \h1743, compared to radio
upper-limits and infrared observational data in the soft state. The
predicted SED in the soft state is consistent with the observational
constraints. In both figures, the optical/infrared (OIR) emission in the soft
state is assumed to originate from the outer parts of the accretion disc
\citep{Coriatetal2009}, therefore we do not attempt to fit the OIR data with our jet 
model.

A detailed modelling of the accretion flow is out of the scope of this work,
however, for illustration purpose we fitted  the OIR to X-ray data shown in
Fig.~\ref{fig:sed_gx339} and \ref{fig:sed_h1743} with the self irradiated
accretion flow model \textsc{diskir} \citep{Gierlinskietal2008,
Gierlinskietal2009}, combined with a Gaussian to model the Fe $K_{\alpha}$ line
and reflection when needed.  The results are shown by the dashed lines in
Fig.~\ref{fig:sed_gx339} and \ref{fig:sed_h1743}. The parameters of the
\textsc{diskir} models are listed in Table~\ref{tab:diskir}. Although these
parameters were obtained using a proper fit procedure leading to a good
statistical representation of the data (reduced $\chi^2$ close to unity), these
are probably not the best fits. In fact, we found considerable model degeneracy
and most of the model parameters are poorly constrained. We did not attempt to
explore the parameter space or make quantitative calculations of the parameter
uncertainties. These models are just shown as example of plausible accretion
flow parameters that give a reasonable description of the data.

Regarding the hard state model, we note that the maximum energy of the radiating
electrons used in D15 ($\gamma_{\rm max}=10^6$) implies that the jet produces
synchrotron radiation up to tens of MeVs. In this case, the predicted jet
synchrotron emission in X-rays is not much below the measured \textit{RXTE}
flux. As the internal shock model also predicts strong variability of the
synchrotron emission (see Malzac 2014, D15), the jet  may contribute
significantly to the X-ray variability. If this was the case, the use of the
X-ray PDS as a tracer of the accretion flow variability would be questionable
and our implementation of the model would not be fully self-consistent. For this
reason, in the present work, we reduced the maximum energy of the electrons to a
much lower value of  $\gamma_{max} = 5\times10^3$.  As  can be seen in the right
panel of Fig.~\ref{fig:sed_gx339}, this change has negligible effects on the
shape and normalization of the radio to IR SED, but implies  the synchrotron
emission is cut-off below the energy range of \textit{RXTE}. This ensures that
the jet does not contribute at all to the observed fast hard X-ray variability
used as model input.  We note that the synchrotron cut-off is not taken into
account in the current version of \ishem.  The exponential cut-off was simply
added to the model a posteriori at an energy determined from $\gamma_{max}$ and
the average magnetic field in the optically thin region.  There is no strong
observational constraints on the location of the synchrotron cut-off in \gx339,
and  we do not exclude  that this cut-off could be located at higher energies.
In this case, the contribution of modelled jet synchrotron to the \textit{RXTE}
band could be strongly reduced by choosing a steeper index for the electron
energy distribution $p$. In the hard state of \gx339, the 1-sigma uncertainties
on the measurement of the optically thin spectral slope in infrared
\citep{Gandhietal2011} would allow for $p$ to be increased up to $p\simeq 3$.
This would remain consistent with acceleration models and would dramatically
reduce the jet  contribution in X-rays, but other parameters of the model, such
as the jet kinetic power or opening angle, would have to be changed in order to
fit the hard state radio-IR SED. Finally, we note that the PDS of the optically
thin jet synchrotron emission predicted by the model is not very different from
the PDS of the input fluctuations (see Malzac 2014), so even if the jet
dominates the X-ray variability, the observed X-ray PDS may remain a reasonable
tracer of the fluctuations in the accretion flow.

\section{Discussion}\label{sec:discussion}

Our results show that there is no need for a dramatic change in the jet
properties in the soft state (such as a decrease of its total kinetic power).
The drop in jet radiative efficiency due to the smaller amplitude of the
fluctuations in the soft state quantitatively accounts for the quenching of the
radio emission and is consistent with the current upper-limits although in the
case of \gx339\ a minor change in the jet geometry is also required. The reader
should note that the non-simultaneity of the X-ray PDS and the radio
observations in the case of \gx339\ could also explain the slight over-prediction
of the model.  Alternatively an increase of the average jet Lorentz factor in
the soft state could also reduce considerably the observed radio flux due to
beaming effects \citep{Maccarone2005}. Of course one cannot exclude that some
other jet properties change at the hard to soft transition. In particular, a
modest decrease of the jet power could also occur and would produce an even
fainter radio emission, but we have performed the present analyses under the
most conservative conditions.

The major point limiting the results is the interpretation of the X-ray PDS
timing information in the soft state. In the hard state, the X-ray emission is
dominated by the Comptonised emission from the hot flow/corona, which is
strongly variable. Whereas in the soft state, a very stable disc component
dominates over that of the corona, and causes the observed X-ray PDS with
weak variability. The variability of the weak non-thermal tail in the soft state
is poorly known but observations of \cygx\ suggest that it is at least as
variable as in the hard state \citep{Churazovetal2001, Gierlinskietal2010}.
The disc contribution to the X-ray PDS
appears to overshadows that of a strongly varying corona.
We found indications that this is the case in \gx339\ and \h1743.  The
quality of the soft state data sets considered in this paper is not good enough
to disentangle accurately the disc and coronal components. But for both sources
we detect a significant increase of rms variability amplitude with energy band.
For instance, in \gx339\ the \mbox{0.07--5 Hz} rms amplitude variability increases
from 1.49 percent in the (disc dominated) 2.5-5.7 keV band to 8.8 percent  in
the (corona dominated)  7-15 keV band.  Unfortunately the low count rate in the
hard band prevents a good determination of the PDS that could be used as input
to our model. However it is obvious that the higher rms  would lead to radio
fluxes that are above the current observed limits for \gx339.

In the framework of our model this implies  that the soft state jet is
driven by the disc rather than the corona. This is not unexpected. Indeed, there
is reasonable evidence suggesting that the hard state corona takes the form of a
hot, optically thin, hard X-ray emitting accretion flow (see e.g. Done et al.
2007) which constitutes the central part of the accretion flow  and probably
drives the jet. In the soft state this hot flow is replaced by a thermally
emitting accretion disc extending down to the last stable orbit. Such disc may
also launch a jet.  The nature  of the soft state corona is not elucidated and
is probably of very different nature from that of the hard state. It is perhaps
unrelated to the jet. For instance it was suggested that the soft state corona
could be made of small scale active magnetic regions above and below the
accretion disc \citep{Zdziarskietal2002}. The weak non-thermal hard X-ray emission of the
soft state appears to have  spectral and timing properties that are very
different from that of the hot flow emission in hard state (Done et al. 2007).
The jets launched from the strongly variable hot accretion flow of the hard
state and the jets  launched from the much steadier accretion disc of the soft
state will have a very different appearance,  even if both kind of jets have
similar properties and  kinetic power.

We  note that the radio quenching may also be related to the change in
variability pattern of the disc, which appears to be strongly variable in the
hard state and very stable in the soft state \citep{Uttleyetal2011,
DeMarcoetal2015}. Alternatively it is possible that the jet variability is
associated only to the band limited X-ray noise observed in the hard state and
not to the flicker noise component of the soft state. In either case, the
predicted radio flux in the soft state would be much lower than what is obtained
here using the full X-ray variability as input to the model. This would make the
detection of soft state jets even more elusive. In any case, our results
illustrate that the radio emission is not a robust tracer of the jet kinetic
power, especially in the soft-state.

Is the presence of dark jets in the soft state of X-ray binaries consistent with
jet launching models? Hot and geometrically thick accretion flows are generally
believed to be more efficient at launching jets than the thin discs of the soft
state \citep{Meier2001, SikoraBegelman2013,Avaraetal2016}. However recent
studies of accretion disc coupled to a jet through the
\citet{BlandfordPayne1982} mechanism indicate that thin discs can eject a larger
fraction of the accretion power than geometrically thick discs
\citep{Petruccietal2010}. In fact, from a theoretical point of view, the thermal
disc-dominated state may harbour an even more powerful jet than the hard state.

Finally we note that a weak jet was reported in the soft state of \cygx\
\citep{Rushtonetal2012}. In this source, the radio flux is often detected in
softer states. The radio emission is correlated with the coronal emission in all
states, from soft to hard, which points to the jet activity being driven by the
hot corona \citep{Zdziarskietal2011}.  However, the soft states of \cygx\ are
characterized by a much stronger coronal component and X-ray variability than
those of the sources, like \gx339\  or \h1743\ where a strong radio quenching is
observed.  In fact, \cygx\ does not seem to reach the canonical soft state.
Besides, the jetted emission of \cygx\ is likely to be more complex than in
classical transient low-mass X-ray binaries because of the interaction of the
jet with the wind of its supergiant companion which may cause both additional
jet dissipation and free-free absorption.

\section{Conclusions}
We have shown that in the context of the internal shock model, the suppression
of the radio emission in soft state is naturally expected even if a jet is still
present and as powerful as in the hard state. Although we do not rule out the
jet quenching paradigm, the possible presence of a powerful dark jet in the
soft state of \xrbs\ should be investigated further.  The compared evolution of
the radio spectrum and X-ray PDS during state transition could bring new
constraints on the evolution of the jet properties between the hard and soft
states. In addition, further investigations are needed to clarify the respective
variability behaviour of the spectral components associated to the disc and
corona in the different spectral state in order to determine which component
drives the jet variability. Moreover, our predictions of radio emission are not
far from the upper-limits currently available. Our results for \gx339 already
indicate that some properties of the jets must change during the transition even
if their power remains the same. Deeper radio observations and future \jwst\
\mbox{mid-IR} spectra can bring more constraints and help determine if indeed a dark
jet is present in \xrb\ sources in the soft state. Finally, the search for
evidence for the interaction of jets with their environment could be a way to
detect dark jets in long-term soft sources.

\section*{Acknowledgements}
The authors thank the anonymous referee for useful comments. J.M. thanks the Institute of Astronomy (Cambridge) for hospitality. S.D. thanks
the Observatoire Midi-Pyr\'{e}n\'{e}es (Tarbes) for hospitality. This work is
part of the CHAOS project ANR-12-BS05-0009 supported by the French Research
National Agency (http://www.chaos-project.fr). P.G. acknowledges support from
STFC (grant reference ST/J003697/2). This paper has made use of up-to-date
SMARTS optical/near-infrared light curves that are available at
www.astro.yale.edu/smarts/xrb/home.php. The Yale SMARTS XRB team is supported by
NSF grants 0407063 and 070707 to Charles Bailyn. This research has made use of a
collection of ISIS functions (ISISscripts) provided by ECAP/Remeis observatory
and MIT (http://www.sternwarte.uni-erlangen.de/isis/).

\bibliographystyle{mn2eMOD}
\bibliography{references}{}

\bsp

\label{lastpage}

\end{document}